\documentclass[aps,nofootinbib,twocolumn,prl,pacs,class-pre]{revtex4}
\usepackage{graphicx}
\usepackage{amsfonts}
\usepackage{amssymb}%

\newcommand{\be}{\begin{equation}}
\newcommand{\ee}{\end{equation}}
\newcommand{\bea}{\begin{eqnarray}}
\newcommand{\eea}{\end{eqnarray}}

\newcommand{\nn}{\nonumber\\}

\begin{document}

\title{Neutrino oscillations in a Robertson-Walker Universe with space time foam}

\author{J. Alexandre}
\address{Kings 's College London, Department of Physics, London WC2R 2LS, U.K.}

\author{K. Farakos}
\address{Department of Physics, National Technical University of Athens, Zografou Campus, 157 80 Athens, Greece}

\author{N. E. Mavromatos}
\address{Kings 's College London,~Department of Physics, London WC2R 2LS, U.K.}

\author{P. Pasipoularides}
\address{Department of Physics,~National Technical University of Athens, Zografou Campus, 157 80 Athens, Greece}

\begin{abstract}
In \cite{AFMP1} we have studied
decoherence models for flavour oscillations in four-dimensional stochastically fluctuating space times
 and discussed briefly the sensitivity of current terrestrial and astrophysical neutrino experiments to such models. 
In this addendum we extend these results to incorporate the effects due to the expansion of the Universe, so that our 
analysis can be useful in studies of extragalactic high-energy neutrinos, such as those coming from Gamma Ray Bursts 
at cosmological distances. Unfortunately for some microscopic models of foam, constructed in the string theory framework, 
we arrive at pessimistic conclusions about the detectability of the decoherence effects via flavour oscillation meaurements.
\end{abstract}

\maketitle

In \cite{AFMP1} we have discussed the propagation of both scalar and Dirac (or Majorana)
flavoured particles in stochastically-fluctuating space times (``foam'')~\cite{stoch}, and we derived the corresponding expressions for the flavour-oscillation probabilities. The presence of a quantum foam results in modifications to the oscillation-probability formula, through damping factors in front of the oscillatory terms and alterations of the oscillation period. Their precise form (as functions of the energy of the probe) and strength depend crucially on the details of the microscopic model used to describe the space-time foamy backgrounds.
The non-observation so far of such effects in various neutrino experiments, both terrestrial and astrophysical, places stringent constraints on the foam parameters, which in turn leads to the exclusion already of some models. The majority, however, of the oscillation models of space time foam, involve damping exponents which are proportional to the square of the square-mass differences
between the various flavours ($(\Delta m^2)^2$), and as such is very suppressed.

Sensitivity to such small effects could only come from cosmic neutrinos, which travel huge distances, of cosmological scale, from their emission at remote celestial objects till observation.
Examples of such cosmic probes of quantum-foam decoherence can be provided by high-energy neutrinos from Gamma-Ray-Bursters (GRB)~\cite{waxman} at high redshifts, $z  \ge 1$.
Currently there are many phenomenological studies of such decoherence effects in
upcoming high-energy cosmic neutrino facilities~\cite{icecube}, however the models used are simplistic
and quite generic models of Lindblad decoherence~\cite{lindblad}, without any attempt to discuss
microscopic situations. Such an attempt was made in \cite{AFMP1}, where the various damping exponents have been derived for some stochastic space-time backgrounds coming from microscopic models of space-time foam, including string-inspired ones~\cite{stoch}. However, there the Universe expansion has been ignored.

In the cases of cosmic probes the effects of the expansion of the Universe cannot be neglected, and hence the analysis of \cite{AFMP1} needs to be extended in order to incorporate them. This is done in this brief note. We shall discuss first the case of scalar probes, deriving the decoherence-induced modifications to the oscillation probability in a Friedman-Robertson-Walker Universe, with space-time foam.
As we shall demonstrate, if one considers the case of a slow-expansion rate for the Universe,
compared to the momentum/energy of the probe, which is a satisfactory approximation for high-energy probes in late eras of the Universe, such as the ones of interest to phenomenology, then the results
for the evaluation of the decoherence exponents for scalar fields can be carried through identical to the Dirac (or Majorana) fermion case.

\vspace{0.5cm}

We commence our discussion by reviewing the different steps leading to the solution of the equation of motion for a scalar field in a
spatially flat Robertson-Walker Universe \cite{birrell}, which is dictated by the astrophysical data and considered here.
Expressed in terms of the cosmic time, the metric is
{\small \be
g_{\mu\nu}=\mbox{diag}(1,-a^2(t),-a^2(t),-a^2(t)),
\ee }
The conformal time $\eta$ is defined as
{\small \be\label{conflat}
g_{\mu\nu}=c(\eta)\eta_{\mu\nu}, ~
c(\eta)=a^2(t),~
d\eta=\frac{dt}{a(t)}~.
\ee }
The curvature scalar is
$R=\frac{3\ddot c}{c^2}-\frac{3}{2c}\left( \frac{\dot c}{c}\right)^2$.

The equation of motion for a scalar field $\phi$ in this background reads,
{\small \be\label{eqmot1}
\frac{1}{\sqrt g}\partial_\mu\left( \sqrt g g^{\mu\nu}\partial_\nu\phi\right)+m^2\phi
=\frac{\dot c}{c}\dot\phi+\ddot\phi-\nabla^2\phi+cm^2\phi=0,
\ee }
where $g=|\mbox{det}(g_{\mu\nu})|$ and
the prime represents derivative with respect to the conformal time $\eta$.
We consider a field depending only on $\eta$ and $x=x^1$ and take the following ansatz
{\small \be\label{ansatz}
\phi(\eta,x)=\frac{\phi_0e^{-ikx}}{\sqrt{c(\eta)}}\chi_k(\eta),
\ee }
where $\phi_0$ is a constant and $k$ is the momentum of the particle.
With this normalization, no ``friction term'' $\dot\chi_k$ is generated from eq.(\ref{eqmot1}), and we obtain
the following equation of motion for $\chi_k$
{\small \be\label{eqmot2}
\ddot\chi_k(\eta)+\omega_k^2(\eta)\chi_k(\eta)=0,
\ee }
where the time-dependent frequency is given by
{\small \be\label{omega}
\omega^2_k(\eta)=k^2+cm^2-\frac{c}{6}R.
\ee }
Note that, in the adiabatic approximation, where $|\dot c|^2<<k^2$ and $|\ddot c|<<k^2$,
particle production by the expanding Universe can be neglected.
This approximation, equivalent to neglecting the curvature $R$, 
is physically reasonable for high-energy probes at late eras of the Universe,
which is of phenomenological interest.
We then look for a solution of eq.(\ref{eqmot2}) of the form
{\small \be\label{ansatzchi}
\chi_k(\eta)=F_k(\eta)\exp\left(i\int_{\eta_0}^\eta d\eta'~W_k(\eta')\right).
\ee }
From eq.(\ref{eqmot2}), we then obtain:
{\small \be\label{WKB1}
\ddot F+2i\dot FW+F(i\dot W-W^2+\omega^2)=0.
\ee }
The latter equation is solved in the framework of the WKB approximation, for which the small parameter is
{\small \be
\epsilon=\frac{\dot\omega_k}{\omega_k^2},
\ee }
In this expansion framework, the various orders are defined by the order of conformal-time derivatives of $F$ and $W$. To lowest order in the WKB approximation, we
neglect $\ddot F$ in eq.(\ref{WKB1}), which leads to:
\begin{itemize}
\item terms without derivatives: $-W^2+\omega^2=0$, such that $W_k(\eta)=\omega_k(\eta)$;
\item terms with one derivative: $2\dot FW+F\dot W=0$, such that $F_k(\eta)=1/\sqrt{\omega_k(\eta)}$.
\end{itemize}
The solution for the initial scalar field to lowest order in the WKB approximation, reads:
{\small \be\label{smoothsolution}
\phi(\eta,x)=\frac{\phi_0e^{-ikx}}{\sqrt{c(\eta)\omega_k(\eta)}}
\exp\left(i\int_{\eta_0}^\eta d\eta'~\omega_k(\eta')\right).
\ee }
where $\eta_0$ denotes the emission time of the probe.
We express the solution (\ref{smoothsolution})
in terms of the redshift $z$, which is related to the scale factor $a(t)$ by the usual relation
{\small \be
a(t)=\sqrt{c(\eta)}=\frac{1}{1+z},
\ee }
where $a(\mbox{now})=1$, when $z=0$. In terms of the Hubble parameter $H(z)=\dot a/a$,
we obtain to lowest order in the WKB approximation,
{\small \be
\phi(z,x)=(1+z)\frac{\phi_0e^{-ikx}}{\sqrt{\omega_k(z)}}\exp\left( -i\int_{z_0}^z dz'\frac{\omega_k(z')}{H(z')}\right) ,
\ee }
where $z_0 \ge z$ (for $\eta_0\le \eta$), with the suffix $0$ here denoting quantities at emission, not to be confused with the (opposite) notation used in astrophysics. In our a notation, the normalization is that at the time of observation $z=0$.

\vspace{0.5cm}

To study the effect of the stochastically fluctuating expanding Universe on flavour oscillations,
we consider for brevity the case of two flavours $|\phi_\alpha>$ ($\alpha=1,2$), related to the energy eigenstates $|f_i>$ ($i=1,2$) by
{\small \bea
&&|\phi_\alpha>=\sum_iU_{\alpha i}|f_i>~~~~~~<\phi_\alpha|=\sum_iU^\star_{\alpha i}<f_i|\nn
&&|f_j>=\sum_\beta U_{\beta j}^\star|\phi_\beta>~~~~~~<f_j|=\sum_\beta U_{\beta j}<\phi_\beta|~.
\eea }
The unitary matrix $U$ corresponds to the tranformation from the flavour eigenstates basis to the
energy eigenstates basis.
The states $|f_i>$ have masses $m_i$ and satisfy the equation of motion (\ref{eqmot1}).
The corresponding solutions for the frequency and
the amplitude are denoted $W_i(\eta)$ and $F_i(\eta)$.
If we assume the following pure state density matrix for the initial condition (no summation over $\alpha$):
{\small \bea\label{initial}
\rho(\eta_0)&=&|\phi_\alpha(\eta_0)><\phi_\alpha(\eta_0)|\nn
&=&\sum_{ij}U_{\alpha i}U^\star_{\alpha j}|f_i(\eta_0)><f_j(\eta_0)|,
\eea }
the time evolution of the energy eigenstates leads then to the following evolution for the normalized density matrix
{\small \bea
\rho(\eta)&=&\sum_{ij}U_{\alpha i}U^\star_{\alpha j}
|f_i(\eta)><f_j(\eta)|\nn
&=&\sum_{ij}U_{\alpha i}U^\star_{\alpha j}
|f_i(\eta_0)><f_j(\eta_0)|\frac{c(\eta_0)}{c(\eta)}\frac{F_i(\eta)F_j(\eta)}{F_i(\eta_0)F_j(\eta_0)}\nn
&&~~~~~~~~~\times\exp\left(i\int_{\eta_0}^\eta d\eta'(W_i(\eta')-W_j(\eta'))\right)\nn
&=& \sum_{\beta\gamma ij}U_{\alpha i}U^\star_{\alpha j}U^\star_{\beta i}U_{\gamma j}
|\phi_\beta(\eta_0)><\phi_\gamma(\eta_0)|
\frac{c(\eta_0)}{c(\eta)}\frac{F_i(\eta)F_j(\eta)}{F_i(\eta_0)F_j(\eta_0)}\nn
&&~~~~~~~~~\times\exp\left(i\int_{\eta_0}^\eta d\eta'(W_i(\eta')-W_j(\eta'))\right).
\eea }As a consequence, starting from the initial flavour $\alpha$,
the transition probability to the flavour $\beta\ne\alpha$ is
{\small \bea\label{Pab}
P_{\alpha\to\beta}&=&\frac{c(\eta_0)}{c(\eta)}\sum_{ij}
U_{\alpha i}U^\star_{\alpha j}U^\star_{\beta i}U_{\beta j}\frac{F_i(\eta)F_j(\eta)}{F_i(\eta_0)F_j(\eta_0)}\nn
&&~~~~~~~~~\times\exp\left(i\int_{\eta_0}^\eta d\eta'(W_i(\eta')-W_j(\eta'))\right),
\eea }
Note that, if $\eta=\eta_0$, we obtain the standard result of flavour oscillations,
{\small \be
P_{\alpha\to\beta}=\sum_{ij}U_{\alpha i}U^\star_{\alpha j}U^\star_{\beta i}U_{\beta j}=0,
\ee }which can be readily verified, for example, with the choice
{\small \be
U=\left(\begin{array}{cc}\cos\theta&-\sin\theta\\\sin\theta&\cos\theta\end{array}\right).
\ee }
Also, for any $\eta$, equality of masses $m_i=m_j$ leads to
{\small \be
P_{\alpha\to\beta}=\frac{c(\eta_0)F^2_i(\eta)}{c(\eta)F^2_i(\eta_0)}
\sum_{ij}U_{\alpha i}U^\star_{\alpha j}U^\star_{\beta i}U_{\beta j}=0.
\ee }

We consider now the effect of a stochastic background by introducing the following modification of the inverse metric,
in terms of the conformal time,
{\small \bea\label{fluctuatingmetric}
g^{\mu\nu}=c^{-1}(\eta)\tilde g^{\mu\nu}~, \quad 
\tilde g^{\mu\nu}=\eta^{\mu\nu}+h^{\mu\nu}~,
\eea }and the stochastic components $h^{\mu\nu}$ are coordinate-independent and are assumed to be small compared to unity. The equation of motion is then
{\small \be\label{equamot7}
\tilde g^{00}\left(\ddot\phi+\frac{\dot c}{c}\dot\phi\right)
+\tilde g^{01}\left(2\partial_x\dot\phi+\frac{\dot c}{c}\partial_x\phi\right)
+\tilde g^{11}\partial^2_x\phi+c(\eta)m^2\phi=0.
\ee } The ansatz (\ref{ansatz}) leads to
{\small \be\label{equachibis}
\tilde g^{00}\ddot\chi_k(\eta)+2ik\tilde g^{01}\dot\chi_k(\eta)+\Omega_k^2(\eta)\chi_k(\eta)=0,
\ee }where
{\small \bea\label{Omega}
\Omega_k^2(\eta)&=&-\tilde g^{11}k^2+c(\eta)m^2-\tilde g^{00}\frac{c}{6}R\nn
&=&\omega_k^2(\eta)-h^{11}k^2-h^{00}\frac{c}{6}R.
\eea } We look for a solution of eq.(\ref{equachibis}) of the form
{\small \be
\chi_k(\eta)=F_k(\eta)\exp\left(i\int_{\eta_0}^\eta d\eta'~W_k(\eta')\right),
\ee}such that $F$ and $W$ satisfy
{\small \be
\tilde g^{00}\left(\ddot F+2i\dot FW+iF\dot W-FW^2\right)+2ik\tilde g^{01}\left(\dot F+iFW\right)+\Omega^2F=0.
\ee} In the framework of the WKB approximation, neglecting $\ddot F$ leads to:
\begin{itemize}
\item terms without derivatives: $-\tilde g^{00}W^2-2k\tilde g^{01}W+\Omega^2=0$, such that
{\small \be
W=-k\frac{\tilde g^{01}}{\tilde g^{00}}+\frac{1}{\tilde g^{00}}\sqrt{k^2(\tilde g^{01})^2+\tilde g^{00}\Omega^2};
\ee}
\item terms with one derivative: $\tilde g^{00}(2\dot FW+F\dot W)+2k\tilde g^{01}\dot F=0$, such that
{\small \be
F=\frac{1}{\sqrt{k\tilde g^{01}+\tilde g^{00}W}}.
\ee}
\end{itemize} If we  consider only the linear terms in the fluctuations $h^{\mu\nu}$, the latter expressions reduce to
{\small \bea\label{WF1storder}
W_k(\eta)&\simeq&\omega_k(\eta)\left(1-\frac{h^{00}}{2}-\frac{h^{11}k^2}{2\omega_k^2(\eta)}\right)-kh^{01} \nn
F_k(\eta)&\simeq&\frac{1}{\sqrt{\omega_k(\eta)}}\left(1-\frac{h^{00}}{4}+\frac{h^{11}k^2}{4\omega_k^2(\eta)}\right)
\eea } The time-dependent part of the probability (\ref{Pab}) is
{\small \be\label{Peta}
P_\eta=\frac{c(\eta_0)}{c(\eta)}\frac{F_i(\eta)F_j(\eta)}{F_i(\eta_0)F_j(\eta_0)}
\exp\left(i\int_{\eta_0}^\eta d\eta'(W_i(\eta')-W_j(\eta'))\right),
\ee}and should be integrated over the random fluctuations of $h^{\mu\nu}$.
For concreteness, in this note we consider a Gaussian distribution for the latter. Since only $h^{00}$ and $h^{11}$ appear in the expression (\ref{Peta}), we define the average
{\small \be
<\cdots>=\frac{1}{\pi\sigma^2}\int dh^{00} dh^{11}(\cdots)
\exp\left(-(h^{00}/\sigma)^2-(h^{11}/\sigma)^2\right) .
\ee} Note that, in order to define the distribution, we consider quadratic terms in $h^{\mu\nu}$ whereas we kept only the first order in $h^{\mu\nu}$ in the result (\ref{WF1storder}), but this is not contradictory if
$\sigma$ is of order $h^{\mu\nu}$. The lowest order of the WKB approximation gives then
{\small \bea
<P_\eta>&=&\frac{c(\eta_0)}{c(\eta)}\exp\left\{i\int_{\eta_0}^\eta d\eta'\big(\omega_i(\eta')-\omega_j(\eta')\big)\right\} \nonumber\\
&&\times\left(1-i\frac{k^4\sigma^2}{4}\Delta_{ij}(\eta)
\int_{\eta_0}^\eta d\eta'\big(1/\omega_i(\eta')-1/\omega_j(\eta')\big)\right)\nn
&&\times\exp\left\{-\frac{\sigma^2}{16}\left(\int_{\eta_0}^\eta d\eta'\big(\omega_i(\eta')-\omega_j(\eta')\big)\right)^2
\right.\nn
&&~~~~\left.-\frac{\sigma^2k^4}{16}
\left(\int_{\eta_0}^\eta d\eta'\big(1/\omega_i(\eta')-1/\omega_j(\eta')\big)\right)^2\right\}, 
\eea }where
{\small \be
\Delta_{ij}(\eta)=\frac{1}{\omega^2_i(\eta_0)}-\frac{1}{\omega^2_i(\eta)}
+\frac{1}{\omega^2_j(\eta_0)}-\frac{1}{\omega^2_j(\eta)}.
\ee} In order to have a more intuitive understanding of this result, we neglect the curvature $R$ and
take $\omega^2\simeq k^2+cm^2$. We make then an expansion in $|m_i-m_j|/k$ and obtain, in terms of the redshift,
{\small \bea\label{Pabaverage}
&& <P_z>=\left(\frac{1+z}{1+z_0}\right)^2\exp\left\{i\frac{m_i^2-m_j^2}{2k^2}I(z)\right\}\\
&&\times\left(1-i\frac{\sigma^2}{16}\frac{m_j^4-m_i^4}{k^4}I(z)
\frac{z_0^2-z^2+2(z_0-z)}{(1+z_0)^2(1+z)^2}\right)\nonumber \\
&&\times\exp\left\{-\frac{\sigma^2}{32}\left(\frac{m_i^2-m_j^2}{k^2}\right)^2I^2(z)\right\} 
\quad\nn
&& \simeq \quad \left(\frac{1+z}{1+z_0}\right)^2 \exp\left\{-\frac{\sigma^2}{32}\left(\frac{m_i^2-m_j^2}{k^2}\right)^2I^2(z)\right\}\nonumber \\
&& \times \exp\left\{i\left(\frac{m_i^2-m_j^2}{2k^2}
- \frac{\sigma^2}{16}\frac{m_j^4-m_i^4}{k^4}
\frac{z_0^2-z^2+2(z_0-z)}{(1+z_0)^2(1+z)^2}\right)I(z)\right\} \nonumber
\eea }
where
{\small \be
I(z)=k\int_{\eta_0}^\eta d\eta'c(\eta')=-k\int_{z_0}^z \frac{dz'}{H(z')}\frac{1}{(1+z')^2}~,
\ee}$z_0$ ($z$) denotes the redshift at emission (observation) of the particle and we have assumed that $\sigma^2 \ll 1$, which, as we shall see later, characterises semi-realistic microscopic models of foam. The oscillation probability is, of course, the real part of the above expression, and thus we observe that the effects of the foam, apart from introducing damping factors, also modify the oscillation period (c.f. imaginary exponents). Notice also that $\sigma^2$ is allowed in general to depend on the energy of the probe, which is found explicitly in some string-inspired models~\cite{emnw,emn2,sarkar}.

In (\ref{Pabaverage}), the first line corresponds to the oscillation term in a smooth expanding Universe;
the second line leads to a correction to the phase of these oscillations, arising from the metric fluctuations,
and the third line represents a damping, also generated by the random fluctuations of the metric.
Note that, for a flat space time, $z=z_0$ and $I(z)=k(t-t_0)$, such that we recover the results of \cite{AFMP1}. Assuming the standard Cosmological-Constant-Dark Matter Model ($\Lambda$CDM),
{\small \be
H(z)=H_0\sqrt{\Omega_\Lambda+(1+z)^3\Omega_M},
\ee}where the various energy density parameters $\Omega_i$ are defined, as usual,
in units of the critical density of the Universe, we thus obtain
for the case of flavoured particles, such as neutrinos, emitted at a period in the Universe with red-shift $z_0$,
{\small \bea\label{cosmo}
I(z)&=&\frac{k}{H_0}\int_z^{z_0} \frac{du}{(1+z')^2}\frac{1}{\sqrt{\Omega_\Lambda+(1+z')^3\Omega_M}}~.
\eea }

To extend the above result to the case of neutrinos (Dirac or Majorana)
we first note that the motion of fermions in a curved space time is described in terms of the vierbeins $e_\mu^\alpha$, which span the
flat tangent space time at each point $x^\mu$ of the manifold, and are related to the metric by
$g_{\mu\nu}=e_\mu^a e_\nu^b\eta_{ab}$, where Latin indices refer to the local inertial (flat, tangent) frame and are contracted with the Minkowski metric $\eta_{ab}$, whereas Greek indices are contracted with the metric $g_{\mu\nu}$. The equation of motion for fermions in a curved space time is \cite{birrell}
{\small \be\label{eqmot3}
 i\gamma^ae_a^\mu\left( \partial_\mu+\Gamma_\mu\right)\psi -m \psi=0,
\ee} where the spin connection is
{\small \be
\Gamma_\mu=\frac{1}{8}[\gamma^a,\gamma^b]e_a^\nu\nabla_\mu e_{b\nu},
\ee} and the gamma matrices are defined in the local inertial frame, and therefore satisfy
{\small \be
\{\gamma^a,\gamma^b\}=2\eta^{ab}.
\ee} With the conformally flat metric (\ref{fluctuatingmetric}), it is easy to check that
{\small \be
e^\mu_a=\frac{1}{\sqrt{c(\eta)}}\left(\delta^\mu_a+\frac{1}{2}h^\mu_a\right)+{\cal O}(h^2).
\ee} If we neglect the time derivatives of the scale factor, the Christoffel symbols vanish, as well as
the spin connections. The equation of motion for fermions, then, reads
{\small \be
i\gamma^a\left(\partial_a+\frac{1}{2}h^\mu_a\partial_\mu\right)\psi-\sqrt{c(\eta)}m\psi=0,
\ee} and a multiplication by the complex conjugate operator leads to
{\small \be\label{eqmot5}
\big(\partial^\rho\partial_\rho+h^{\mu\nu}\partial_\mu\partial_\nu+c(\eta)m^2\big)\psi
=\big(\tilde g^{\mu\nu}\partial_\mu\partial_\nu+c(\eta)m^2\big)\psi=0.
\ee}
This equation is similar to eq.(\ref{equamot7}) for a scalar field, when the derivatives of the scale factor are neglected when compared with the momenta/energies of the probe, and hence in this approximation of slow expansion, the previous results for the decoherence damping exponents of the scalar field carry through intact to the fermion case (above we have discussed for concreteness Dirac fermions, but the extension to the Majorana case is straightforward and does not alter the final result).

\vspace{0.5cm}

Let us now make a few remarks on the result (\ref{Pabaverage}), (\ref{cosmo}). As charactertistic of the Gaussian space-time fluctuations~\cite{stoch,AFMP1}, the damping exponent depends on the square of the time traversed by the probe, i.e. the damping terms are of the form ${\rm exp}\left(-D~t^2\right)$.
Should one use other distributions for the space-time fluctuations,  for instance the Cauchy-Lorentz distribution~\cite{AFMP1}, that could arise in some string inspired models of quantum foam~\cite{stoch},
then the decoherence damping assumes a Lindblad exponential form~\cite{lindblad}.
From (\ref{Pabaverage}), the decoherence-induced damping exponent is of the form:
{\small \be \label{decexp}
{\rm exponent} = -\frac{\sigma^2}{32}\left(\frac{m_i^2-m_j^2}{k^2}\right)^2 k^2
\left(\int_{z_0}^z \frac{dz'}{H(z')}\frac{1}{(1+z')^2}\right)^2~.
\ee}
Some remarks are in order concerning the nature of the variance $\sigma^2$. Its precise estimate
requires microscopic models of the stochastically fluctuating space time.

One such model is inspired from string theory, namely the D-particle foam model~\cite{emnw}, in which
a space time fluctuation arises from topologically non-trivial interactions of open strings
(representing matter excitations on a brane world scenario) with D-particle defects.
The scenario involves capture and re-emission of the open string by the defect, which is microscopically
represented by the appearance of intermediate string states stretched between the D-particle and the brane world. This leads~\cite{emn2} to \emph{causality}-respectful time delays, proportional to the
energy $E$ of the incident string,
{\small \be
\Delta t \sim \alpha ' E = \frac{E}{M_s^2},
\label{dtime}
\ee}
On the other hand, the recoil of the D-particle defect during the above process implies local distortions of the space time, which are of Finsler type, in the sense of depending on both the coordinates and the momentum transfer $\Delta {\vec k}$ of the string state. In particular, one obtains non diagonal metrics, which for small recoil velocities correspond to deviations from the Minkowski metric of the form
{\small \be
      h_{0i} \sim u_i~, ~i = 1,2, 3 \qquad u_i = \frac{g_s}{M_s}\Delta k_i
\ee}
As discussed in detail in \cite{sarkar}, quantum fluctuations of the recoil velocity arise by summing up over world-sheet genera, which translate in turn to stochastic quantum fluctuations in the induced finsler space time. The corresponding distributions are Gaussian, with a variance $\sigma^2$, which, over the time scale of the interaction of the D-particle with the string state $\delta t  $ (\ref{dtime}), becomes
{\small \be
   \sigma^2_{ij}  = \langle u_i u_j \rangle = \sigma^2 \delta_{ij} = g_s^2\frac{(\Delta t)^2}{\alpha'}\delta_{ij}
   = g_s^2 \frac{E^2}{M_s^2}\delta_{ij}~.
   \label{variance}
\ee}
Using (\ref{variance}) in (\ref{decexp}), then, and approximating to leading order in small quantities $E \sim |\vec k|$ (which for the case of high-energy neutrinos is a perfectly good approximation), we
obtain for the decoherence damping exponent the estimate:
{\small \bea\label{expo2}
&& {\rm exponent} = -\frac{g_s^2}{32M_s^2H_0^2}\left(m_i^2-m_j^2\right)^2 \times \nonumber \\
&&\left(\int_{z_0}^0 \frac{dz'}{\sqrt{\Omega_\Lambda + (1 + z')^3\Omega_M }}\frac{1}{(1+z')^2}\right)^2~,
\eea}for the $\Lambda$CDM cosmology, where we set the red-shift of observations $z=0$.

We first observe that for this model of space-time foam, the decoherence damping exponent is independent of the probe's energy or momenta.
Taking into account that the value of Hubble parameter today $H_0 \sim 10^{-24}M_P$, where $M_P \sim 10^{19}$~GeV is the four dimensional Planck mass scale, we can express (\ref{expo2}) as:
$ 10^{48}\left(\frac{M_P}{M_s}\right)^2 \frac{(\Delta m^2)^2}{M_P^4} $.
For $g_sM_P/M_s \sim 1$, which is a phenomenologically realistic situation, allowing also for large string scales and very weak string couplings, and taking into account that the $k$-independent red-shift factor
in the last line of (\ref{expo2}) is of order one for red-shifts $z_0 \le 5$ we are interested in here,
one observes that the resulting decoherence coefficient is unobservably small ($< 10^{-72}$), for phenomenologically realistic values of mass-squared differences between neutrino flavours of order $\Delta m^2 \sim 10^{-3}-10^{-5}$ eV$^2$.
In fact to obtain decoherence exponents of order one, one needs probes that have travelled the age of the observable Universe. This has to do with the fact that the factor
$\frac{(\Delta m^2)^2}{M_P^4}$ is numerically of the same order as the (observed)  Cosmological constant
$10^{-122}M_P^4$ ! In a similar way the other decoherence corrections in (\ref{Pabaverage}) are also unobservably small for this model. Thus, this model of D-particle foam cannot be falsified via decoherence damping tests.

On the other hand, for other distributions of recoil velocities, such as the Cauchy-Lorentz (CL) distribution, with characteristic function $\Phi (\xi) = \frac{\gamma}{\pi (\xi^2 + \gamma^2)}$,
used in \cite{AFMP1}, one has exponents of the form:
{\small \be
{\rm exponent}_{\rm CL} \sim \gamma H_0^{-1}\frac{\Delta m^2}{2k} \left(I(z)H_0/k\right)
\ee}that is, exponents of Lindblad type, varying linearly with the (cosmic) time.
Again the $k$-independent red-shift time factor $I(z)H_0/k$ (c.f. (\ref{cosmo})) is contributing terms of order one for the region of red-shifts
of interest $z \le 5$.
The CL type distributions may be encountered~\cite{AFMP1} in populations of D-particle defects, as opposed to the virtual Gaussian quantum fluctuations we have discussed previously.

Since the parameter $\gamma$ is essentially arbitrary in such models, one might hope that the CL distributions may lead to observable effects for cosmological distances $1 < z \le 5$, for infrared neutrinos. In fact, for $\delta m^2 \in \{10^{-3} - 10^{-5} \}$ eV$^2 = 10^{-59} - 10^{-61}~M_P^2$, $H_0 \sim 10^{-24}~M_P$ we obtain the following estimate for the decoherence damping exponent of the CL distributions, modulo the
order-one $I(z)H_0/k$ factors:
$\frac{\gamma M_P}{2k} (10^{-35} - 10^{-37})$~. This
is of order one for neutrinos of eV energies, if $\gamma \sim 10^7$ for the lower bound, which however is an unnatural value for a dispersion on velocities. For values $\gamma \le 1$, becomes of order $10^{-4}$ for the lower bound, only for
infrared neutrinos of energies in the meV or less.

\section*{Aknowledgements}

The work of N.E.M. is partially supported by the European Union
through the Marie Curie Research and Training Network \emph{UniverseNet}
(MRTN-2006-035863).


\begin{thebibliography}{99}


\bibitem{AFMP1}  J.~Alexandre, K.~Farakos, N.~E.~Mavromatos and P.~Pasipoularides
   Phys.\ Rev.\  D {\bf 77} (2008) 105001.

\bibitem{stoch} L.~H.~Ford,
  Phys.\ Rev.\  D {\bf 51}, 1692 (1995);
  Int.\ J.\ Theor.\ Phys.\  {\bf 38} (1999) 2941 and references therein;
J.~R.~Ellis, N.~E.~Mavromatos and D.~V.~Nanopoulos,
  Gen.\ Rel.\ Grav.\  {\bf 32}, 127 (2000);
N.~E.~Mavromatos and Sarben~Sarkar,
  Phys.\ Rev.\  D {\bf 74}, 036007 (2006);
  New J.\ Phys.\  {\bf 10}, 073009 (2008).

\bibitem{waxman} E.~Waxman and J.~N.~Bahcall,
  Phys.\ Rev.\ Lett.\  {\bf 78}, 2292 (1997);
  Astrophys.\ J.\  {\bf 541}, 707 (2000).


\bibitem{icecube} L.~A.~Anchordoqui, H.~Goldberg, M.~C.~Gonzalez-Garcia, F.~Halzen, D.~Hooper, Subir~Sarkar and T.~J.~Weiler,
  Phys.\ Rev.\  D {\bf 72}, 065019 (2005).
   D.~Hooper, D.~Morgan and E.~Winstanley,
  Phys.\ Lett.\  B {\bf 609}, 206 (2005).

\bibitem{lindblad} G.~Lindblad,
  Commun.\ Math.\ Phys.\  {\bf 48}, 119 (1976);
  F.~Benatti and R.~Floreanini,
  Phys.\ Rev.\  D {\bf 64}, 085015 (2001);
For latest results see: G.~L.~Fogli, E.~Lisi, A.~Marrone,
D.~Montanino and A.~Palazzo,
arXiv:0704.2568 [hep-ph].


\bibitem{birrell} N.~D.~Birrell and P.~C.~W.~Davies,
{\it  Cambridge, Uk: Univ. Pr. ( 1982) 340p}

\bibitem{emnw} J.~R.~Ellis, N.~E.~Mavromatos, D.~V.~Nanopoulos and M.~Westmuckett,
  Int.\ J.\ Mod.\ Phys.\  A {\bf 21}, 1379-1444 (2006);
J.~R.~Ellis, N.~E.~Mavromatos and M.~Westmuckett,
  Phys.\ Rev.\  D {\bf 71}, 106006(1-13) (2005)

\bibitem{emn2} J.~R.~Ellis, N.~E.~Mavromatos and D.~V.~Nanopoulos,
  Phys.\ Lett.\  B {\bf 665}, 412-417 (2008) and \emph{addendum}
  in [arXiv:0901.4052 [astro-ph.HE]], Phys. Lett. B in press.



\bibitem{sarkar} N.~E.~Mavromatos and S.~Sarkar,
  arXiv:0812.3952 [hep-th];
N.~E.~Mavromatos and R.~J.~Szabo,
  Phys.\ Rev.\  D {\bf 59}, 104018 (1999).


\end{thebibliography}
\end{document}